\documentclass[twocolumn,showpacs,preprintnumbers,
amsmath,amssymb,aps,prd,nofootinbib,superscriptaddress,
eqsecnum]{revtex4}
\usepackage{graphicx,color}

\makeatletter
\renewcommand{\p@subsection}{}
\makeatother

\newcommand{\Slash}[1]{\ooalign{\hfil/\hfil\crcr$#1$}}

\begin{document}

\title{Fate of charmed mesons near chiral symmetry restoration in hot matter}

\author{Chihiro Sasaki}
\affiliation{%
Frankfurt Institute for Advanced Studies,
D-60438 Frankfurt am Main,
Germany
}
\affiliation{%
Institute of Theoretical Physics, University of Wroclaw, 
PL-50204 Wroclaw, 
Poland
}

\date{\today}

\begin{abstract}
Chiral thermodynamics of charmed mesons is formulated at finite temperature
within a $2+1+1$-flavored effective Lagrangian incorporating heavy quark 
symmetry. The charmed-meson mean fields act as an extra source which breaks
the chiral symmetry explicitly. This leads to effective interactions between
the light and heavy-light mesons, which intrinsically depend on temperature.
Effective masses of the scalar and pseudoscalar charmed-mesons tend to 
approach each other as increasing temperature, so that the splitting 
between the chiral partners is reduced. These chiral splittings are shown
to be less sensitive to the light-quark flavors, attributed to the 
underlying heavy quark symmetry. Consequently, chiral symmetry restoration is
more manifest for the strange charmed-mesons than for the strange light
mesons. The effective masses are also compared with the results in the 
one-loop chiral perturbation theory. A substantial difference is found at a 
relatively low temperature, $T \sim f_\pi$.
\end{abstract}


\maketitle

%%%%%%%%%%%%%%%%%%%%%%%%%%%%%%%%%%%%%%%%%%%%%%%%%%%%
\section{Introduction}
\label{sec:int}
%%%%%%%%%%%%%%%%%%%%%%%%%%%%%%%%%%%%%%%%%%%%%%%%%%%

Heavy flavors, charm and beauty, are produced at the initial stage of
the high-energy heavy-ion collisions, so that they are expected to carry
the dynamical history of a created matter, the Quark-Gluon Plasma (QGP)
(see, e.g.~\cite{RvH} for a review). Recent experimental observations have
revealed that charm quarks are thermalized~\cite{phenix,star,alice1,alice2},
contrary to earlier anticipation. Charge fluctuations calculated in lattice 
QCD also indicate that the charmed mesons are deconfined together with 
light-flavor mesons~\cite{latcharm}. Given those observations, comprehensive
exploration for the chiral aspects of the heavy-light hadrons increases its
importance. Their characteristics in a hot system are important input to 
disentangle the transport properties of hadronic matter and QGP.

In-medium modifications of the charmed mesons have been extensively
studied by using QCD sum rules~\cite{qsr1,qsr2,qsr3} and effective 
theories~\cite{su4,su42,cc,su43,njl,MGOT,YS}. In constructing effective
Lagrangians for the heavy-light mesons, besides spontaneous chiral symmetry
breaking, heavy quark symmetry is a vital ingredient~\cite{HQS}.
The pseudo-scalar $D$ and vector $D^\ast$ states fill in the same multiplet $H$,
forming the lowest spin partners. Their low-energy dynamics is dominated by 
interactions with Nambu-Goldstone (NG) bosons, pions~\cite{Georgi,Wise,BD,YCCLLY}.
Introducing the multiplet including $D$ and $D^\ast$ inevitably accompanies
another multiplet $G$ which contains a scalar $D_0^\ast$ and axial-vector $D_1$
states. Those parity partners, $H$ and $G$, become degenerate when the chiral
symmetry is restored~\cite{NRZ,BH}. However, it is not a priori obvious how
the chiral mass splitting, which is $\sim 350$ MeV in matter-free space, is 
resolved: one increases and another decreases in their masses, or the two of 
them decrease/increase. This crucially affects e.g. the dissociation of the
charmed mesons in matter.

Aside from the chiral $SU(4)$ approach where the charm sector suffers from
a huge explicit breaking of the extended flavor symmetry, a self-consistent 
study for the thermal charmed-mesons with implementing heavy quark symmetry has 
received little attention. In the present paper, we formulate a chiral effective
theory for the light and heavy-light mesons in a hot/dense medium under the 
mean field approximation. A special attention is devoted to the in-medium masses
of the chiral partners. We discuss the role of the heavy quark symmetry in 
the thermal evolution of the chiral mass splittings for the non-strange and 
strange states. The result is also compared with the one-loop self-energy 
calculated in chiral effective field theory.

%%%%%%%%%%%%%%%%%%%%%%%%%%%%%%%%%%%%%%%%%%%%%%%%%%%%
\section{Effective Lagrangian}
\label{sec:eff}
%%%%%%%%%%%%%%%%%%%%%%%%%%%%%%%%%%%%%%%%%%%%%%%%%%%

To describe the light-quark sector, we take the standard linear sigma
model Lagrangian with three flavors:
\begin{eqnarray}
{\mathcal L}_{\rm L}
&=&
\bar{q}\left(i\Slash{\partial} - g T^a\left(
\sigma^a + i\gamma_5\pi^a
\right)\right)q
\nonumber\\
&&
{}+ {\rm tr}\left[\partial_\mu\Sigma^\dagger\cdot\partial^\mu\Sigma\right]
{}- V_{\rm L}(\Sigma)\,,
\end{eqnarray}
where the potential, including $U(1)_A$ breaking effects, is
\begin{eqnarray}
V_{\rm L}
&=&
m^2{\rm tr}\left[\Sigma^\dagger\Sigma\right]
{}+ \lambda_1\left({\rm tr}\left[\Sigma^\dagger\Sigma\right]\right)^2
\nonumber\\
&&
{}+\lambda_2{\rm tr}\left[\left(\Sigma^\dagger\Sigma\right)^2\right]
{}- c\left(\det\Sigma + \det\Sigma^\dagger\right)
\nonumber\\
&&
{}- {\rm tr}\left[h\left(\Sigma + \Sigma^\dagger\right)\right]\,,
\end{eqnarray}
with the chiral field $\Sigma = T^a\Sigma^a
= T^a\left(\sigma^a + i\pi^a\right)$ as a $3 \times 3$ complex matrix
in terms of the scalar $\sigma^a$ and the pseudoscalar $\pi^a$ states.
The last term with $h = T^a h^a$ breaks the chiral symmetry explicitly.

Heavy-light meson fields with negative and positive parity are
introduced as~\cite{NRZ,BH}~\footnote{
 We will use the notations $D$ for a scalar and $P$ for a pseudo-scalar
 state, unless otherwise stated with quantum numbers.
}
\begin{eqnarray}
H
&=&
\frac{1 + \Slash{v}}{2}\left[
P_\mu^\ast\gamma^\mu + iP\gamma_5
\right]\,,
\\
G
&=&
\frac{1 + \Slash{v}}{2}\left[
-iD_\mu^\ast\gamma^\mu\gamma_5 + D
\right]\,,
\label{GH}
\end{eqnarray}
and chiral eigenstates are given from those parity eigenstates via
\begin{equation}
{\mathcal H}_{L,R}
=
\frac{1}{\sqrt{2}}\left(G \pm iH\gamma_5\right)\,.
\end{equation}
The field operators are transformed under the chiral and heavy quark
symmetries as
\begin{eqnarray}
{\mathcal H}_{L,R} 
&\to& 
S{\mathcal H}_{L,R}g_{L,R}^\dagger\,,
\\
\Sigma 
&\to& 
g_L\Sigma g_R^\dagger\,,
\end{eqnarray}
with the group elements $g_{L,R} \in SU(3)_{L,R}$ 
and $S \in SU(2)_{Q=c}$.

The lowest-order Lagrangian to first oder in $\Sigma$ and to
zeroth order in $1/m_Q$ up to two heavy-light fields is~\cite{NRZ,BH} 
\begin{eqnarray}
{\mathcal L}_{\rm HL}^{\rm kin}
&=&
\frac{1}{2}{\rm Tr}\left[
\bar{\mathcal H}_L iv\cdot\partial{\mathcal H}_L
{}+ \bar{\mathcal H}_R iv\cdot\partial{\mathcal H}_R
\right]\,,
\\
V_{\rm HL}^{(2)}
&=&
\frac{m_0}{2}{\rm Tr}\left[\bar{\mathcal H}_L{\mathcal H}_L 
{}+ \bar{\mathcal H}_R{\mathcal H}_R\right]
\nonumber\\
&&
{}+ \frac{g_\pi}{4}{\rm Tr}\left[
\Sigma^\dagger\bar{\mathcal H}_L{\mathcal H}_R
{}+ \Sigma\bar{\mathcal H}_R{\mathcal H}_L
\right]
\nonumber\\
&&
{}- i\frac{g_A}{2F_\pi}{\rm Tr}\left[
\gamma_5\Slash{\partial}\Sigma^\dagger\cdot\bar{\mathcal H}_L{\mathcal H}_R
{}- \gamma_5\Slash{\partial}\Sigma\cdot\bar{\mathcal H}_R{\mathcal H}_L
\right]\,,
\nonumber\\
\end{eqnarray}
where traces are taken over Dirac and light-flavor indices.

A self-consistent calculation for the ${\mathcal H}_{L,R}$
requires a further contribution beyond $V_{\rm HL}^{(2)}$.
Minimal extension for this is to add terms including four heavy-light fields.
Following the given transformation properties, one finds in the same order
\begin{eqnarray}
V_{\rm HL}^{(4,0)}
&=&
c_1{\rm Tr}\left[
\bar{\mathcal H}_L{\mathcal H}_L\bar{\mathcal H}_L{\mathcal H}_L
{}+ \bar{\mathcal H}_R{\mathcal H}_R\bar{\mathcal H}_R{\mathcal H}_R
\right]
\nonumber\\
&&
{}+ c_2\left(\left({\rm Tr}\left[\bar{\mathcal H}_L{\mathcal H}_L\right]
\right)^2
{}+ \left({\rm Tr}\left[\bar{\mathcal H}_R{\mathcal H}_R\right]
\right)^2\right)
\nonumber\\
&&
{}+ c_3{\rm Tr}\left[
\bar{\mathcal H}_L{\mathcal H}_R\bar{\mathcal H}_R{\mathcal H}_L
\right]
\nonumber\\
&&
{}+ c_4{\rm Tr}\left[\bar{\mathcal H}_L{\mathcal H}_L\right]
\left[\bar{\mathcal H}_R{\mathcal H}_R\right]\,,
\\
V_{\rm HL}^{(4,1)}
&=&
\kappa_1{\rm Tr}\left[
\Sigma^\dagger\bar{\mathcal H}_L{\mathcal H}_R
\bar{\mathcal H}_R{\mathcal H}_R
{}+ \Sigma\bar{\mathcal H}_R{\mathcal H}_L
\bar{\mathcal H}_L{\mathcal H}_L
\right]
\nonumber\\
&&
{}+ \kappa_2{\rm Tr}\left[
\Sigma^\dagger\bar{\mathcal H}_L{\mathcal H}_L
\bar{\mathcal H}_L{\mathcal H}_R
{}+ \Sigma\bar{\mathcal H}_R{\mathcal H}_R
\bar{\mathcal H}_R{\mathcal H}_L
\right]
\nonumber\\
&&
{}+ \kappa_3\left(
{\rm Tr}\left[\Sigma^\dagger\bar{\mathcal H}_L{\mathcal H}_R\right]
{\rm Tr}\left[\bar{\mathcal H}_R{\mathcal H}_R\right]
\right.
\nonumber\\
&&
\quad
\left.
{}+ {\rm Tr}\left[\Sigma\bar{\mathcal H}_R{\mathcal H}_L\right]
{\rm Tr}\left[\bar{\mathcal H}_L{\mathcal H}_L\right]
\right)
\nonumber\\
&&
{}+ \kappa_4\left(
{\rm Tr}\left[\Sigma^\dagger\bar{\mathcal H}_L{\mathcal H}_R\right]
{\rm Tr}\left[\bar{\mathcal H}_L{\mathcal H}_L\right]
\right.
\nonumber\\
&&
\quad
\left.
{}+ {\rm Tr}\left[\Sigma\bar{\mathcal H}_R{\mathcal H}_L\right]
{\rm Tr}\left[\bar{\mathcal H}_R{\mathcal H}_R\right]
\right)\,.
\end{eqnarray}
Explicit $SU(3)$ symmetry breaking is encoded in the potential with
the following replacement,
\begin{equation}
\Sigma \to 
\lambda h\Sigma\,,
\end{equation}
with $\lambda$ denoting a set of constant parameters.
The entire potential is thus 
\begin{equation}
V_{\rm HL}
=
V_{\rm HL}^{(2)} + V_{\rm HL}^{(4,0)} + V_{\rm HL}^{(4,1)}
{}+ V_{\rm HL}^{\rm exp}\,.
\label{HLpot}
\end{equation}

For our thermodynamic calculations, we employ the mean field approximation.
The SU(2) isospin violation is also neglected, which leads to $\sigma_0$ and
$\sigma_8$ as non-vanishing condensates. Those fields contain both non-strange 
and strange components.
Thus, it is convenient to transpose them into pure non-strange and strange
parts, via
\begin{equation}
\begin{pmatrix}
\sigma_q\\
\sigma_s
\end{pmatrix}
=
\frac{1}{\sqrt{3}}
\begin{pmatrix}
\sqrt{2} & 1 \\
1 & -\sqrt{2}
\end{pmatrix}
\begin{pmatrix}
\sigma_0 \\
\sigma_8
\end{pmatrix}\,.
\end{equation}
The effective quark masses in this base are
\begin{equation}
M_q 
=
\frac{g}{2}\sigma_q\,,
\quad
M_s = \frac{g}{\sqrt{2}}\sigma_s\,.
\end{equation}
The partially conserved axial current (PCAC) hypothesis relates
$\sigma_{q,s}$ with the weak decay constants for pions and kaons:
\begin{equation}
\langle\sigma_q\rangle = f_\pi\,,
\quad
\langle\sigma_s\rangle 
= \frac{1}{\sqrt{2}}\left(2f_K - f_\pi\right)\,.
\end{equation}

The scalar charmed-mesons are accommodated in the multiplets,
\begin{equation}
D = (D_q, D_q, D_s)\,,
\end{equation}
where due to the isospin symmetry $D_u=D_d=D_q$.
In-medium masses are calculated from the potential via
\begin{equation}
\Delta M_{D_i} 
=
M_{D_i} - m_c
= \frac{1}{2}\frac{\partial^2 V_{\rm HL}}{\partial D_i^2}\,,
\end{equation}
with $i=u,d,s$ and the charm quark mass $m_c$.
From Eq.~(\ref{HLpot}), one finds
\begin{eqnarray}
\Delta M_D(0^+)
&=&
m_0 + \frac{1}{4}g_\pi^q\sigma_q 
{}+ 2k_0\left(4D_q^2 + D_s^2\right) 
\nonumber\\
&&
{}+ 6k_q\sigma_q D_q^2\,,
\\
\Delta M_{D_s}(0^+)
&=&
m_0 + \frac{1}{2\sqrt{2}}g_\pi^s\sigma_s 
{}+ 2k_0\left(2D_q^2 + 3D_s^2\right)
\nonumber\\
&& 
{}+ 6\sqrt{2}k_s\sigma_s D_s^2\,,
\\
\Delta M_D(0^-)
&=&
m_0 - \frac{1}{4}g_\pi^q\sigma_q 
{}+ 2k_0\left(4D_q^2 + D_s^2\right)\,,
\\
\Delta M_{D_s}(0^-)
&=&
m_0 - \frac{1}{2\sqrt{2}}g_\pi^s\sigma_s 
{}+ 2k_0\left(2D_q^2 + 3D_s^2\right)\,,
\nonumber\\
\label{mass}
\end{eqnarray}
where $k_0, g_\pi^q, g_\pi^s, k_q, k_s$ are functions of the parameters
$c$'s, $\kappa$'s and $\lambda$'s.

In the mean field approximation,
thermodynamics of this system is described by the following potential:
\begin{eqnarray}
\Omega
&=&
\Omega_q + V_{\rm L} + V_{\rm HL}\,,
\\
\Omega_q
&=&
6\,T\sum_{f=u,d,s}
\int\frac{d^3p}{(2\pi)^3}
\left[\ln\left(1-n_f\right)
{}+ \ln\left(1-\bar{n}_f\right)\right]\,,
\nonumber\\
\end{eqnarray}
with the Fermi-Dirac distribution functions
$n_f, \bar{n}_f = 1/\left(1 + e^{(E_f \mp \mu_f)/T}\right)$
and the quasi-quark energies $E_f = \sqrt{p^2 + M_f^2}$.
By minimizing the thermodynamic potential, the four mesonic mean-fields
are determined self-consistently at a given $T$ and $\mu_f$ via
\begin{equation}
\frac{\partial\Omega}{\partial\sigma_q}
= \frac{\partial\Omega}{\partial\sigma_s}
= \frac{\partial\Omega}{\partial D_q}
= \frac{\partial\Omega}{\partial D_s}
= 0\,.
\label{gap}
\end{equation}

In the thermal model applied in heavy-ion collisions, temperature $T$ and 
baryon chemical potential $\mu_B$ are independent parameters. On the
other hand, the strange and charm chemical potentials are fixed via
strange and charm number conservation~\cite{BMS}.
The chemical potential for a particle $i$ is introduced as
\begin{equation}
\mu_i = \mu_B B_i + \mu_s S_i + \mu_c C_i\,,
\end{equation}
with the baryon number $B_i$, strangeness $S_i$ and charm $C_i$ quantum
numbers. In term of the particle number density,
\begin{equation}
n_i = \frac{g_i}{2\pi^2}\int_0^\infty dp\,
\frac{p^2}{e^{(E-\mu_i)/T} \pm 1}\,,
\end{equation}
with the degeneracy factor $g_i$, the conservation conditions are give as
\begin{equation}
V \sum_i n_i S_i = 0\,,
\quad
V \sum_i n_i C_i = 0\,,
\end{equation}
where $V$ denotes the volume of a system.
In the following, we will take $\mu_B=0$, so that the
strange and charm conservation are fulfilled trivially, $\mu_s=\mu_c=0$.

%%%%%%%%%%%%%%%%%%%%%%%%%%%%%%%%%%%%%%%%%%%%%%%%%%%%%%%%%%%%%%%
\section{Chiral thermodynamics in the mean-field approximation}
\label{sec:mf}
%%%%%%%%%%%%%%%%%%%%%%%%%%%%%%%%%%%%%%%%%%%%%%%%%%%%%%%%%%%%%%%%

Conventionally model parameters are fixed at zero temperature
so as to reproduce the meson masses and decay constants.
However, regarding effective models as an approximation of QCD
in low energy, the parameters in the effective Lagrangian vary
with temperature. Such intrinsic thermal effects are carried by 
the higher-lying hadrons, and formally introduced via functional 
integration in deriving Green's functions. At finite temperature,
a reliable way to extract those effective parameters is to match
them with the observables in lattice QCD.

In the light-flavor sector, the sigma meson mass $m_\sigma$ is often
treated as an adjustable parameter. In the vacuum the nature of
the lowest-lying scalar state remains not fully understood yet due to
a strong mixing to other states with the same quantum number, e.g.
a tetra-quark state~\cite{PDG}. Thus, we use $m_\sigma$ as a parameter
to adjust the pseudo-critical temperature of chiral symmetry
restoration to the lattice result, $T_{\rm pc}=154$ MeV extracted from the 
chiral susceptibility $\partial\langle\bar{q}q\rangle/\partial m_q$~\cite{LQCD}. 
The resultant choice in the linear sigma model is $m_\sigma=400$ MeV. 
We will use other parameters fixed in the vacuum~\cite{SW}, summarized 
in Table~\ref{paraL}.
%%%%%%%%%%%%%%%%%%%%%%%%%%%%%%%%%%%%%%%%%%%%%%
\begin{table*}
\begin{center}
\begin{tabular*}{12cm}{@{\extracolsep{\fill}}ccccccc}
\hline
$c$ [GeV] & $m$ [GeV] & $\lambda_1$ & $\lambda_2$ & $h_q$ [GeV$^3$]
& $h_s$ [GeV$^3$] & $g$ \\
\hline
$4.81$ & $0.495$ & $-5.90$ & $46.48$ & $(0.121)^3$ & $(0.336)^3$
& $6.5$ \\
\hline
\end{tabular*}
\end{center}
\caption{
Set of parameters in the light sector with 
$m_\sigma = 400$ MeV~\cite{SW}.
}
\label{paraL}
\end{table*}
%%%%%%%%%%%%%%%%%%%%%%%%%%%%%%%%%%%%%%%%%%%%%%%%%

The parameters in the heavy-light sector at zero temperature are
determined as in Table~\ref{paraHL} where the following input was used;
$m_c=1.27$ GeV, $M_D(0^-)=1.868$ GeV, $M_{D_s}(0^-)=1.969$ GeV, 
$M_{D_s}(0^+)=2.318$ GeV, and the pion and kaon decay constants 
$f_\pi=92.4$ MeV, $f_K=113$ MeV~\cite{PDG}.
%%%%%%%%%%%%%%%%%%%%%%%%%%%%%%%%%%%%%%%%%%%%%%
\begin{table*}
\begin{center}
\begin{tabular*}{12cm}{@{\extracolsep{\fill}}cccccc}
\hline
$m_0$ [GeV] & $g_\pi^q$ & $g_\pi^s$ & $k_0$ [1/GeV$^2$] 
& $k_q$ [1/GeV$^3$] & $k_s$ [1/GeV$^3$] \\
\hline
$1.04$ & $3.78$ & $2.61$ & $-(1/0.74)^2$ & $-(1/0.44)^3$
& $-(1/0.53)^3$ \\
\hline
\end{tabular*}
\end{center}
\caption{
Set of parameters in the heavy-light sector.
}
\label{paraHL}
\end{table*}
%%%%%%%%%%%%%%%%%%%%%%%%%%%%%%%%%%%%%%%%%%%%%%%%%
The current experimental value for the $D(0^+)$ mass by the Particle
Data Group has a rather large error, $M_D^{\rm PDG}(0^+)=2318 \pm 29$ MeV.
With the given parameters, the model predicts $M_D(0^+)/M_D^{\rm PDG}(0^+)
= 0.96$. The coupling constant $g_\pi^q$ is extracted also from the decay
mode, $D(0^+) \to D(0^-)\pi$, yielding $g_\pi^{\rm PDG}=3.6 
= 0.95 \times g_\pi^q$.

\begin{figure}
\begin{center}
\includegraphics[width=8cm]{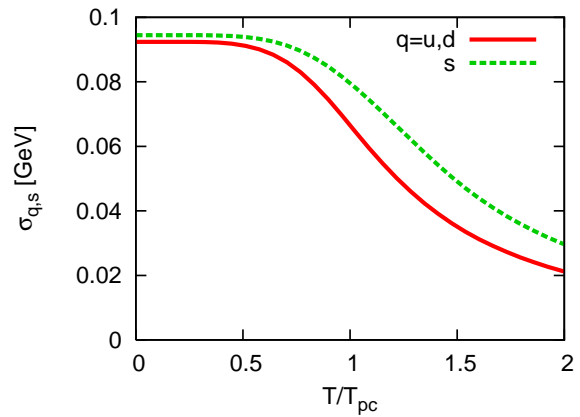}
\caption{
Thermal expectation values of the mean fields, 
$\sigma_q$ and $\sigma_s$. The pseudo-critical temperature fixed
from the chiral susceptibility is $T_{\rm pc} = 154$ MeV.
}
\label{fig:mf}
\end{center}
\end{figure}

In Fig.~\ref{fig:mf} we show thermal expectation values of the $\sigma_q$ and 
$\sigma_s$ in the mean field approximation.
%%%%%%%%%%%%%%%%%%%%%%%%%%%%%%%%%%%%%%%%%%%%
%%%%%%%%%%%%%%%%%%%%%%%%%%%%%%%%%%%%%%%%%%%%%%%%%%
One readily sees that the chiral crossover is extremely broad; at $T_{\rm pc}$
the order parameter $\sigma_q$ reaches $\sim 0.73\times\sigma_q(T=0)$, 
whereas in lattice QCD the bilinear quark condensate drops more rapidly 
to almost a half of its vacuum value~\cite{LQCD}. Also, the $\sigma_s$ is
accompanied to a large extent by the non-strange condensate $\sigma_q$, which
is again inconsistent with the lattice observation.

This is traced back to a rather strong mixing between the light and
heavy-light mean fields. The charmed-meson mean fields act as an extra 
source which breaks the chiral symmetry explicitly. The effective symmetry 
breaking is induced by
\begin{eqnarray}
h_q^\ast
&=& 
h_q - D_q^2\left(\frac{1}{2}g_\pi^q + 2k_qD_q^2\right)\,,
\\
h_s^\ast
&=& 
h_s - \frac{1}{\sqrt{2}}D_s^2\left(\frac{1}{2}g_\pi^s + 2k_sD_s^2\right)\,.
\end{eqnarray}
At finite temperature, the size of $h_{q,s}^\ast$ varies via coupled 
equations of motion. As clearly indicated in Fig.~\ref{fig:mf}, this 
results in the light-flavor $SU(2+1)$ symmetry promoted to approximate 
$SU(3)$ toward the chiral crossover, i.e. $h_q^\ast \simeq h_s^\ast$.

Since the charm quark mass is much larger than characteristic temperatures
associated with the chiral crossover, such a strong modification of the
$\sigma_{q,s}$ by the charmed mesons is unrealistic. In fact, there is no
evidence found in lattice QCD that the $N_f=2+1$ thermodynamics is affected
by the dynamical charm quark around the chiral crossover~\cite{charmLQCD}.
The difference starts to appear above $T \sim 300$ MeV where the charmed
mesons are supposed to be resolved already~\cite{latcharm}. 
The present mean-field framework deals with the charmed mesons as static 
background fields which mix with the light scalar mean fields. This transmutes
a light-flavor insensitivity to the condensates $\sigma_{q,s}$, which is an 
artifact of the current setup. Therefore, the 
sigma mean fields need to be adjusted by reducing the coupling constant(s) 
effectively, in such a way that the chiral properties are fit to the lattice 
observation. 

\begin{figure}
\begin{center}
\includegraphics[width=8cm]{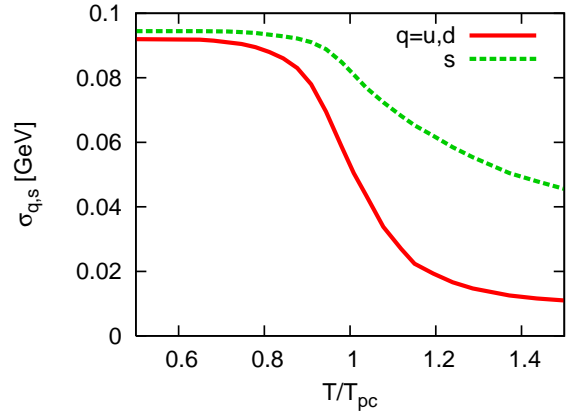}
\caption{
Assumed profiles of thermal expectation values of 
$\sigma_q$ and $\sigma_s$ with $T_{\rm pc} = 154$ MeV.
}
\label{fig:corr}
\end{center}
\end{figure}

\begin{figure*}
\begin{center}
\includegraphics[width=8cm]{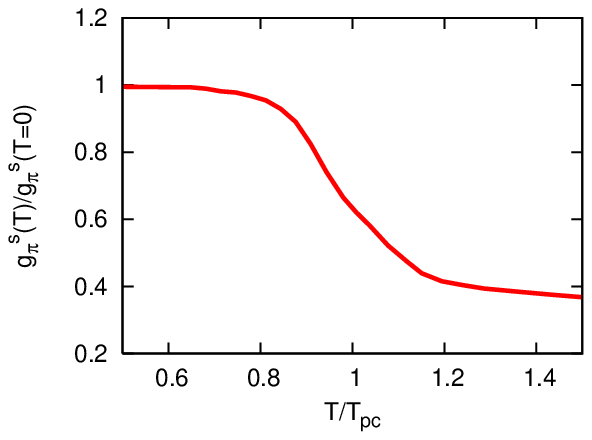}
\includegraphics[width=8cm]{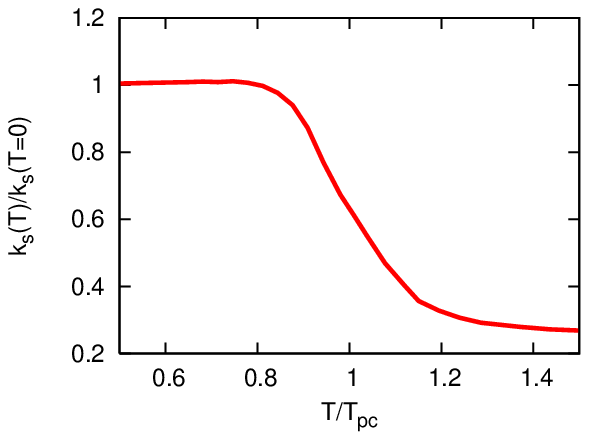}
\caption{
Effective interactions among light and heavy-light mesons. 
}
\label{fig:tdep}
\end{center}
\end{figure*}

\begin{figure}
\begin{center}
\includegraphics[width=8cm]{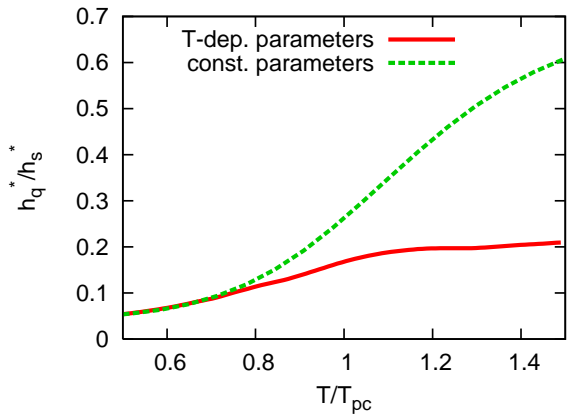}
\caption{
Induced symmetry breaking with the effective couplings $g_\pi^s(T)$ and
$k_s(T)$ (solid) and that with the constant interactions (dashed).
}
\label{fig:mixT}
\end{center}
\end{figure}

In the light-flavor sector, a hierarchy lying among up, down and
strange flavors is spoiled since the heavy quark symmetry over-influences
the light-flavored quark condensates. Therefore, the correct tendency
needs to be restored by controlling the effective interaction between 
the light and heavy-light mesons. Practically, we need to constrain
the size of $h^\ast_q/h^\ast_s$ within a certain range.
To this end, we dictate in-medium changes to the parameters in 
the Lagrangian with the use of a reliable set of $\sigma_{q,s}(T)$.
Possible profiles for the sigma fields are as given in Fig.~\ref{fig:corr},
which are consistent with the quark condensates calculated in the $N_f=2+1$
lattice QCD~\cite{LQCD}.
%%%%%%%%%%%%%%%%%%%%%%%%%%%%%%%%%%%%%%%%%%%%
%%%%%%%%%%%%%%%%%%%%%%%%%%%%%%%%%%%%%%%%%%%%%%%%%%
In order to control a relative size of $\sigma_q$ to $\sigma_s$,
we will introduce thermal modifications into the parameters
in the strange sector, $g_\pi^s$ and $k_s$, whereas the non-strange parameters
are kept to be the values fixed at $T=0$~\footnote{
 Alternatively, one can use $g_\pi^s$ and $k_s$ being their vacuum values and
 introduce $g_\pi^q(T)$ and $k_q(T)$. This does not alter our main conclusion.
}. 
The effective couplings $g_\pi^s(T)$
and $k_s(T)$ are determined by solving the gap equations~(\ref{gap}) with the 
given profiles for the $\sigma_{q,s}$. In Fig.~\ref{fig:tdep}, one finds that 
those interactions decrease their strengths as the system approaches the chiral
symmetry restoration, so that the overall symmetry breaking $h_s^\ast$ is not 
reduced significantly.
%%%%%%%%%%%%%%%%%%%%%%%%%%%%%%%%%%%%%%%%%%%%
%%%%%%%%%%%%%%%%%%%%%%%%%%%%%%%%%%%%%%%%%%%%%%%%%%
This is shown in Fig.~\ref{fig:mixT}.
%%%%%%%%%%%%%%%%%%%%%%%%%%%%%%%%%%%%%%%%%%%%
%%%%%%%%%%%%%%%%%%%%%%%%%%%%%%%%%%%%%%%%%%%%%%%%%%

\begin{figure*}
\begin{center}
\includegraphics[width=8cm]{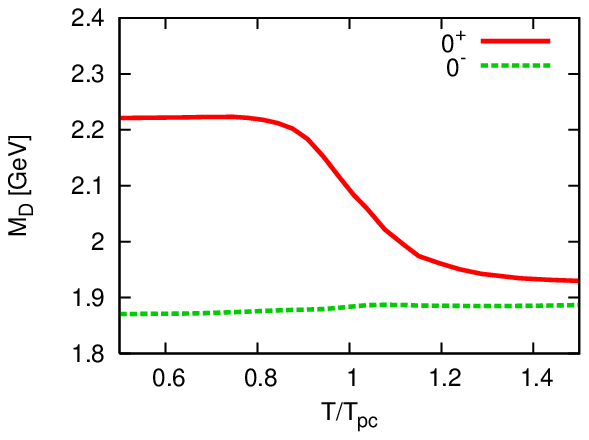}
\includegraphics[width=8cm]{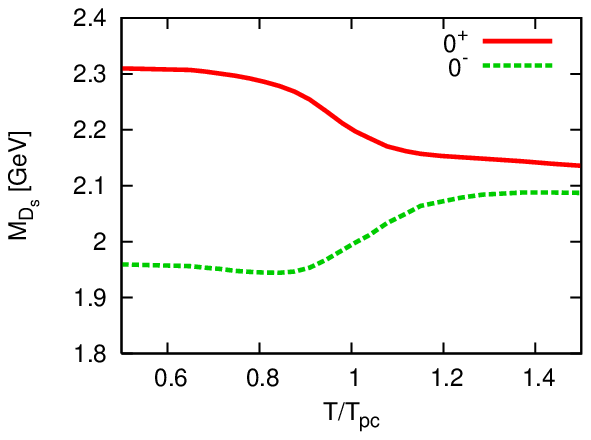}
\caption{
In-medium masses of the non-strange (left) and strange (right)
charmed-mesons with positive and negative parity.
}
\label{fig:md}
\end{center}
\end{figure*}

The charmed-meson masses~(\ref{mass}) are now calculated consistently to the 
lattice result in the light-flavor sector, as summarized in Fig.~\ref{fig:md}.
%%%%%%%%%%%%%%%%%%%%%%%%%%%%%%%%%%%%%%%%%%%%
%%%%%%%%%%%%%%%%%%%%%%%%%%%%%%%%%%%%%%%%%%%%%%%%%%
The parity partners approach each other as temperature is increased
both in the non-strange and strange sector, in consistent with the chiral
restoration. The two pseudo-scalar states have the same trend that their
masses are increasing with temperature, although the non-strange meson mass
exhibits a rather weak modification. On the other hand, the two scalar states
drop significantly; the non-strange meson mass by $\sim 200$ MeV and the strange
meson mass by $\sim 100$ MeV. The mass splittings between the non-strange
and strange states are around $200$ MeV above $T_{\rm pc}$ due to the fact
that the chiral symmetry in the strange sector is not restored yet.
Nevertheless, the chiral mass splittings between the scalar and pseudo-scalar
states are almost of the same size,
\begin{equation}
\delta M_D(T_{\rm pc}) \sim \delta M_{D_s}(T_{\rm pc}) \sim 200\,\mbox{MeV}\,,
\label{HLchiral}
\end{equation}
i.e. {\it the chiral mass differences in the heavy-light sector are blind to 
the light flavors.} This is a striking difference from the chiral properties
of the light mesons, and is attributed to the heavy quark symmetry possessed 
by the leading-order Lagrangian in $1/m_Q$ expansion.
In contrast, the chiral SU(4) model, where the charmed mesons are treated on the 
equal footing to the non-strange and strange mesons, yields a qualitatively
different result from Eq.~(\ref{HLchiral}); $\delta M_D$ is much smaller than 
$\delta M_{D_s}$, similar to the light meson masses~\cite{su4}.

The effective coupling $g_\pi^s(T)$ also affects the hadronic decays involving
the $D_s$ states. Those decay modes violate the isospin symmetry and
thus they are suppressed. This is followed dominantly via the $\pi^0$-$\eta$
mixing~\cite{BEH}. The decay width of $D_s(0^+) \to D_s(0^-) + \pi^0$ is
\begin{equation}
\Gamma \simeq \frac{\left(g_\pi^s\right)^2}{4\pi}\bar{p}_\pi
\delta_{\pi^0\eta}^2\,,
\end{equation}
where $\bar{p}_\pi = |\vec{p}_\pi|$ is the three-momentum of the pion
in the rest frame of the decaying particle, and the $\pi^0$-$\eta$
mixing is given by
\begin{equation}
\delta_{\pi^0\eta}
= \frac{2m_\pi^2(m_u-m_d)}{(m_\eta^2-m_\pi^2)(m_u+m_d)}\,.
\end{equation}
We recall that the $g_\pi^s$ decreases with increasing temperature as given 
in Fig.~\ref{fig:tdep}, i.e. the $D_s$ meson tends to be decoupled from the
light-flavor sector. On top of the small isospin breaking, this decay mode 
becomes more suppressed in approaching $T_{\rm pc}$, due to the $g_\pi^s(T)$
which controls the onset of chiral symmetry restoration. The decay process, 
$D_s(1^+) \to D_s(1^-) + \pi^0$, is quenched as well.

%%%%%%%%%%%%%%%%%%%%%%%%%%%%%%%%%%%%%%%%%%%%%%%%%%%%
\section{Results in chiral perturbation theory}
\label{sec:chpt}
%%%%%%%%%%%%%%%%%%%%%%%%%%%%%%%%%%%%%%%%%%%%%%%%%%%

At zero temperature, the composition of the lowest scalar meson is nontrivial
because of a strong mixing between the conventional quarkonium and tetra-quark
states. The scalar state around 1 GeV is a good candidate of the lowest 
$\bar{q}q$-dominated meson.
Thus, as long as a characteristic temperature is lower than the chiral crossover,
the sigma meson in the linear sigma model can be integrated out since it lies 
well above the pion mass scale. 
The resultant Lagrangian contains only the NG bosons and
the chiral symmetry is non-linearly realized. The basic building block is 
the 1-form $\alpha_\perp$,
\begin{equation}
\alpha_\perp^\mu
=
\frac{1}{2i}\left(\partial^\mu\xi\cdot\xi^\dagger
{}- \partial^\mu\xi^\dagger\cdot\xi\right)\,,
\end{equation}
where the NG bosons $\pi$ are embedded in $\xi$ as $\xi = e^{i\pi/f_\pi}$.
The interaction to the heavy-light meson fields~(\ref{GH}) is quantified
by the following Lagrangian~\cite{HRS}:
\begin{eqnarray}
{\mathcal L}_{\rm int}
&=&
k\left(
{\rm Tr}\left[H\gamma_\mu\gamma_5\alpha_{\perp}^\mu\bar{H}\right]
{}+ {\rm Tr}\left[G\gamma_\mu\gamma_5\alpha_{\perp}^\mu\bar{G}\right]
\right.
\nonumber\\
&&
\left.
{}- i{\rm Tr}\left[G\alpha_{\perp\mu}\gamma^\mu\gamma_5\bar{H}\right]
{}+ i{\rm Tr}\left[H\alpha_{\perp\mu}\gamma^\mu\gamma_5\bar{G}\right]
\right)\,,
\nonumber
\\
\end{eqnarray}
with the coupling constant $k=0.59$ extracted form the decay $D^\ast \to D\pi$.
At finite temperature, thermal corrections to the charmed-meson masses are
induced from the self-energy. The major temperature dependence is carried by
the NG bosons. Therefore, we approximate the in-medium propagator of the D mesons
to their vacuum forms. The pion propagator is replaced as
\begin{equation}
\frac{1}{p^2 - m_\pi^2 + i\epsilon}
\to
\frac{1}{p^2 - m_\pi^2 + i\epsilon}
{}- \frac{2\pi i}{e^{|p_0|/T}-1}\delta(p^2-m_\pi^2)\,.
\end{equation}

Substituting those propagators to the self-energy at one loop~\cite{HRS},
one obtains the following thermal corrections to the scalar ($\Pi_S^\ast$)
and the pseudo-scalar ($\Pi_P^\ast$) states:
\begin{eqnarray}
&&
\Pi_{S}^\ast
=
C_2(N_f)
\frac{k^2}{f_\pi^2}\left[
\int\frac{d^3p}{(2\pi)^3}n_\pi(m_\pi,T)
\right.
\nonumber\\
&&
\left.
{}- \left(M_P + M_S + \frac{m_\pi^2}{M_S}\right)A_0^\ast(m_\pi,T)
\right.
\nonumber\\
&&
\left.
{}- \frac{M_S^2 - m_\pi^2}{M_P}A_S^\ast(m_\pi,T)
{}- M_P A_P^\ast(m_\pi,T)
\right]\,,
\\
&&
\Pi_{P}^\ast
=
C_2(N_f)
\frac{k^2}{f_\pi^2}\left[
\int\frac{d^3p}{(2\pi)^3}n_\pi(m_\pi,T)
\right.
\nonumber\\
&&
\left.
{}- \left(M_P + M_S + \frac{m_\pi^2}{M_P}\right)A_0^\ast(m_\pi,T)
\right.
\nonumber\\
&&
\left.
{}- \frac{M_P^2 - m_\pi^2}{M_P}A_P^\ast(m_\pi,T)
{}- M_S A_S^\ast(m_\pi,T)
\right]\,,
\end{eqnarray}
where $M_{S,P}$ represents the scalar (pseudo-scalar) charmed meson mass,
and the functions $A_0^\ast$ and $A_{S,P}^\ast$ are introduced in terms of
the Bose-Einstein distribution function $n_\pi$ as
\begin{eqnarray}
n_\pi
&=&
\frac{1}{e^{\omega/T}-1}\,,
\quad
\omega = \sqrt{p^2 + m_\pi^2}\,,
\\
A_0^\ast
&=&
\int\frac{d^3p}{(2\pi)^3}\frac{n_\pi(\omega)}{\omega}\,,
\\
A_{S,P}^\ast
&=&
\int\frac{d^3p}{(2\pi)^3}\frac{n_\pi(\omega)}{\omega + M_{S,P}}\,.
\end{eqnarray}
The group factor $C_2(N_f)$ is defined as 
$(T^a)_{ij}(T^a)_{jl} = C_2(N_f)\delta_{il}$.
The chiral mass difference is thus
\begin{eqnarray}
\delta M
&=& 
C_2(N_f)
\frac{k^2}{f_\pi^2}m_\pi^2\left[
\left(\frac{1}{M_P} - \frac{1}{M_S}\right)A_0^\ast(m_\pi)
\right.
\nonumber\\
&&
\left.
{}+ \frac{1}{M_S}A_S^\ast(m_\pi)
{}- \frac{1}{M_P}A_P^\ast(m_\pi)
\right]\,,
\end{eqnarray}
which is less sensitive to a temperature rise.
If the mass parameters $M_{S,P}$ carry a certain temperature dependence
so that $M_S$ approaches $M_P$ as temperature is increased, the chiral
partners tend to be degenerate. Such non-trivial intrinsic thermal
effects must show up beyond the standard one-loop perturbative method.

\begin{figure}
\begin{center}
\includegraphics[width=8cm]{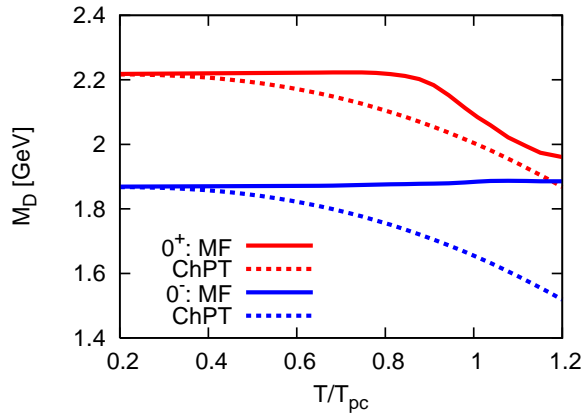}
\caption{
Comparison of the non-strange charmed-meson masses in the present mean-field
model (solid) with the corresponding results in the one-loop chiral perturbation
theory (dashed).
}
\label{fig:chpt}
\end{center}
\end{figure}

In Fig.~\ref{fig:chpt}, the results are compared with the effective masses
of the non-strange charmed mesons discussed in the previous section.
%%%%%%%%%%%%%%%%%%%%%%%%%%%%%%%%%%%%%%%%%%%%
%%%%%%%%%%%%%%%%%%%%%%%%%%%%%%%%%%%%%%%%%%%%%%%%%%
The pion loop yields a monotonic decrease with temperature for the scalar
and pseudo-scalar states. A difference from the mean-field theory starts to
appear at rather low temperature, $T \sim f_\pi \sim 0.67 m_\pi$.
This temperature cannot be fixed solely by the 
symmetries, but relies on the dynamics of the theories, in particular the 
onset of chiral criticality. This is absent in the standard chiral perturbation
theory, whereas present in the mean-field theory considered in the previous
section via a self-consistent prescription. 
The agreement with the self-consistent result will be expected at a higher 
temperature when higher loops and/or more resonances are included.

%%%%%%%%%%%%%%%%%%%%%%%%%%%%%%%%%%%%%%%%%%%%%%%%%%%%%%%
\section{Conclusions}
\label{sec:conc}
%%%%%%%%%%%%%%%%%%%%%%%%%%%%%%%%%%%%%%%%%%%%%%%%%%%%%

In this paper we have formulated a chiral mean-field theory for the light and
heavy-light mesons at finite temperature based on the heavy quark symmetry.
In order to avoid an unrealistically strong mixing between the light-flavor
and the charmed meson sector, effective interactions depending on temperature
are introduced, which can be extracted from the chiral condensates calculated
in lattice QCD. The coupling of the strange charmed meson to the sigma meson,
$g_\pi^s$, becomes quenched as temperature is increased toward the chiral 
pseudo-critical point $T_{\rm pc}$. Our main result is that the chiral mass 
splittings are essentially insensitive to the light-quark flavors, in spite 
of a non-negligible explicit breaking of the chiral $SU(3)$ symmetry. 
This ``blindness'' of the charm quark to the light degrees of freedom is 
dictated by the heavy quark symmetry. In contrast,
the kaon and its chiral partner masses become degenerate at a higher temperature 
than $T_{\rm pc}$, indicating a delay of the $SU(3)$ symmetry 
restoration. In the heavy-light sector, on the other hand, the strange charmed
meson captures the onset of chiral symmetry restoration more strongly than 
the strange light meson does. 
The quenched $g_\pi^s$ leads also to a strong suppression of the scalar $D_s$ 
decay toward $T_{\rm pc}$, on top of the suppression due to the small isospin 
violation. The same should be carried over to the $B$ and
$B_s$ mesons with which the heavy quark symmetry is more reliable.

Although the present model does not enable to handle a confinement/deconfinement 
transition, it reliably captures the chiral aspects of the charmed mesons 
constrained by the heavy flavor symmetry.
Given the lattice QCD observations that non-strange and strange hadrons
seem to be resolved to their constituents around the chiral crossover,
the system contains low-lying mesons and the chiral fermions (quarks) 
when the baryon chemical potential is sufficiently small.
The main result shown in this work will be robust in hadronic phase up to 
the chiral restoration point as long as there is no strong first-order transition.

Those medium modifications may yield some consequences on the nuclear 
modification factor and the elliptic flow in high-energy heavy-ion collisions,
on top of the effects discussed in \cite{HFR} where the scalar $D$ and $D_s$
states are chirally unmodified. Scenarios of charmonium suppression~\cite{RBC}
will certainly be affected by the modified heavy-light mesons around the chiral
symmetry restoration. In particular, the dissociation of the charmed mesons
needs to be reexamined.

Application of our approach to a dense system requires further implementation
of (i) strange and charm number conservation, and (ii) reliable constraint(s) 
on the effective 
interaction between the light and heavy-light mesons. At zero density, the latter
has been provided by the lattice chiral condensates. Alternatively, a more 
microscopic framework will enable us to derive the in-medium coupling as
a function of temperature and density. This becomes an essential input
especially for the study of various charge fluctuations~\cite{SRfluc}.

%%%%%%%%%%%%%%%%%%%%%%%%%%%%%%%%%%%%%%%%%%%%%%%%%%%%%%%%
\subsection*{Acknowledgments}
%%%%%%%%%%%%%%%%%%%%%%%%%%%%%%%%%%%%%%%%%%%%%%%%%%%%%%

We acknowledge stimulating discussions with K.~Redlich.
This work has been partly supported by the Hessian LOEWE initiative
through the Helmholtz International Center for FAIR (HIC for FAIR),
and by the Polish Science Foundation (NCN) under
Maestro grant 2013/10/A/ST2/00106.

%%%%%%%%%%%%%%%%%%%%%%%%%%%%%%%%%%%%%%%%%%%%%%%%%%%%%%%%
%%%%%%%%%%%%%%%%%%%%%%%%%%%%%%%%%%%%%%%%%%%%%%%%%%%%%%%%%

\end{document}